\documentclass[11pt]{article}
\usepackage{graphicx}

\begin{document}

\title{BCS as Foundation and Inspiration: The Transmutation of Symmetry}

\author{Frank Wilczek \\
\small\it Center for Theoretical Physics\\[-1ex]
\small \it Department of Physics\\[-1ex]
\small \it Massachusetts Institute of Technology\\[-1ex]
\small\it Cambridge, MA 02139}
\date{}

\maketitle

\begin{abstract}
The BCS theory injected two powerful ideas into the collective consciousness of theoretical physics: pairing and spontaneous symmetry breaking.  In the 50 years since the seminal work of Bardeen, Cooper, and Schrieffer, those ideas have found important use in areas quite remote from the stem application to metallic superconductivity.   This is a brief and eclectic sketch of some highlights, emphasizing relatively recent developments in QCD and in the theory of quantum statistics, and including a few thoughts about future directions.   A common theme is the importance of symmetry {\it transmutation},  as opposed to the simple {\it breaking\/} of electromagnetic $U(1)$ symmetry in classic metallic superconductors.
\end{abstract}

\newpage
\frenchspacing

The Bardeen-Cooper-Schrieffer (BCS) theory of superconductivity \cite{bcs} has been fruitful in many ways.   Most obviously, of course, it provided profound and at many points surprising concrete insights into the superconducting state of solids, right from the start.   It predicted for instance the very different effect of the onset of superconductivity on acoustic versus electromagnetic relaxation, due to the different signs in coherence factors, which is a delicate quantum-mechanical effect.   And it provided the intellectual foundation for such wonders as the Josephson effects and Andreev reflection.     

The influence of BCS theory on the broader discipline of theoretical physics has been no less profound.  Two key ideas abstracted from BCS theory, that have been widely transplanted and borne abundant fruit, are {\it pairing\/}  and {\it dynamical symmetry breaking}.   Pairing was an essentially new idea, introduced by Cooper and brought to fruition by BCS.   The symmetry breaking aspect was mostly implicit in the original BCS work, and in earlier ideas of Fritz London and Landau-Ginzburg; but the depth and success of the BCS theory seized the imagination the theoretical physics community, and catalyzed an intellectual ferment.    The concept of spontaneous symmetry breaking was promptly made explicit, generalized, and put to use by several physicists including Anderson, Josephson, Nambu, and Goldstone.  The flexibility and transformative power of these ideas revealed itself gradually, in applications to phenomena that at first sight appear to have little or nothing in common with superconductivity.   

From a wealth of possible material, I've chosen to discuss some relatively recent developments close to my own work, that I think well illustrate how naturally the basic BCS concepts combine with other ideas of fundamental and emergent symmetry, often with dramatic consequences.    A common theme is that symmetry {\it breaking\/} forms a special case of a more general phenomenon: symmetry {\it transmutation}.    

\section{QCD Meets BCS: Color-Flavor Locking, Confinement, and Chiral Symmetry Breaking}

Quantum chromodynamics or QCD, having run the gauntlet of many exquisite quantitative confrontations with observation, is now established as the fundamental theory of the strong interaction \cite{qcdReview}.    QCD is a challenging theory to understand, however, and not primarily because of its technical complexity\footnote{``Technical complexity'' is a time-dependent concept.  I've heard graduate students accustomed to a diet of high supersymmetry, Calabi-Yau manifolds and intersecting D-branes refer to QCD as ``trivial'', with no evident ironic intent.}.   The real challenge comes in relating the wonderfully ``trivial'' basic equations to observed reality.   The primary ingredients of QCD are massless gluons and nearly massless quarks ($u, d, s$; the heavy quarks $c, b, t$ are a separate, and much easier, study).   The observed hadrons, of course, are neither massless like gluons nor fractionally charged like quarks.   Many techniques have been deployed to bridge the chasm separating theory world and physical world, but in my opinion none is clearer nor more elegant than the straightforward application of BCS ideas to the regime of high density.   (Here by ``high density'' I intend large baryon number density, at low temperature.)  \cite{densityReview}

\bigskip
\subsection{QCD Meets BCS}
\bigskip

A wise principle states ``It is more blessed to ask forgiveness than permission.''  In that spirit, we consider the possibility of constructing a description of high-density QCD based on its elementary degrees of freedom, quarks and gluons.  

At first sight this approach looks extremely promising.  High density means large fermi surfaces.  Neglecting interactions, the low-energy excitations are associated with modes near the fermi surface: a mode just above the fermi surface, empty in the ground state, becomes occupied, or a mode just below becomes empty.    Since the fermi surface is large, all modes near the fermi surface carry large momentum and energy.    Thus scattering among the low-energy excitations will either involve only small angles, and leave the distribution of particles over modes nearly unchanged, or else bring in large momentum transfers, and therefore weak coupling (asymptotic freedom).   It appears, therefore, that perturbation theory should be a good approximation.

But when one actually does the calculations, one finds infrared divergences.   They arise from two sources:
\begin{itemize}
\item The preceding argument only concerns the quarks.   Its central point is that Pauli blocking removes the infrared divergences that usually arise through low-virtuality quarks.   Gluons, however, are not subject to any such effect.   Color electric forces are screened by the quark medium, but color magnetic forces remain long-ranged, and lead to infrared divergences.
\item Interacting fermions are subject to the Cooper instability.   One has many near-zero energy excitations at zero momentum, associated with particle-particle or hole-hole pairs carrying equal and opposite three-momenta $\pm \vec {\rm p}$.  Thus in perturbing around the many-body state which is the ground state of the non-interacting theory, that is the fully occupied fermi sphere, one is engaging in highly degenerate perturbation theory.   As a general matter, degenerate perturbation theory can result in significant restructuring of the ground state.    In this specific context, Bardeen, Cooper, and Schrieffer (BCS) taught us that even a small attractive interaction {\it will\/} lead to a drastic re-arrangement of the ground state, by inducing pairing and superfluidity.
\end{itemize}

In conventional superconductors it is quite subtle to find an effective attractive interaction between electrons.   The primary interaction between electrons is the Coulomb interaction, and it is of course repulsive.  To find an attractive interaction one must bring in phonons, retardation, and screening, and concentrate on modes within a thin shell around the Fermi surface.    For many ``unconventional'' superconductors, famously including the cuprates, the mechanism of attraction remains unclear at present, despite much effort to identify it.  But in all known cases the superconducting transition temperature (which reflects the attractive dynamics) is far below the melting temperature (which reflects the primary dynamics).

In QCD the situation is more straightforward, because the primary interaction -- the QCD analogue of the Coulomb interaction -- can already be attractive.   Two separated quarks, each in the triplet {\bf 3} representation, can be brought together in the antisymmetric $\bar{\bf 3}$.   The disturbance in the gluon field due to color charge is then half what it was before.  Since the energy has decreased, the force is attractive. Nothing like this can happen when there's just one type of charge, of course.  The existence of three different color charges is crucial here.   (On the other hand, with larger numbers of colors antisymmetrization yields relatively less reduction in flux, and so the attractive force is relatively weaker.)    

By zeroing the spin -- that is, once again, choosing the antisymmetric channel -- we also remove the sources of magnetic disturbance.    Thus on very general grounds we expect a powerful attractive interaction between quarks in the channel where both colors and spins are antisymmetric.    This intuition is borne out by calculations using one-gluon exchange, instanton models, and direct numerical simulations, though those simulations could and should be sharpened.     

Thus color superconductivity occurs straightforwardly, and should be robust physically, at high density.    What does it mean?   Here is a list of physical effects we can anticipate, by translating  intuition from superconductivity into the language of particle physics:
\begin{itemize}
\item Gluons acquire mass -- that is a way to state the equations of the Meissner effect.   If {\it all\/} the gluons acquire mass,  their exchange will no longer produce infrared divergences.
\item Quarks acquire mass -- that is a way to state the equations of the energy gap.  If {\it all\/} the quarks acquire mass, Cooper's infrared divergence will be removed.  
\item Thus we construct a new ground state, around which our weak-coupling expansion works.  
\item This ground state does not contain massless gluons nor exhibit long-range forces.  In that sense, it exhibits confinement.   We also have the classic phenomenon of confinement -- that is, absence of fractional {\it electric\/} charge in the spectrum, as I'll explain shortly.
\item The energy gap for quarks suggests that chiral symmetry, which is associated with massless quarks, may be broken.   
\end{itemize}
In short, we have the prospect of a phase that exhibits the main non-perturbative features of QCD -- confinement and chiral symmetry breaking -- in a transparent, fully controlled theoretical framework.    Let me emphasize that here I am speaking of a phase of QCD itself, not of some idealization of a model of a caricature of QCD.   Now let's discuss how it's realized, more concretely.   

\bigskip
\subsection {Color-Flavor Locking}
\bigskip

For concreteness, and because my emphasis here is on QCD rather than astrophysics, I will assume as the initial default that all quarks are massless, that they are subject to a common chemical potential, and that electromagnetism can be treated as a perturbation.   I'll circle back to revisit these assumptions, and ask your forgiveness, in due course.   

Because the most attractive channel for quarks is antisymmetric both in color and spin, Fermi statistics requires another source of antisymmetry.    One possibility is antisymmetry in the spatial wave-function of the quark pairs.   For example, we might have $p$-wave pairing.   But for simple, purely attractive interaction potentials, $s$-wave tends to be favored, because it allows pairs from all directions over the Fermi surface to act in phase.  So $s$-wave pairing, if possible, is likely to be favorable.

The remaining possible source of antisymmetry is flavor.   Thus we must pair off {\it different\/} flavors of quarks to take best advantage of the attractive interaction between quarks.   This requirement brings in some significant complications.   Obviously, it means that the one-flavor case is not representative, and that we cannot build up the analysis one flavor at a time.   The two-flavor case also does not go smoothly.    Antisymmetry in flavor and spin (and lack of orbital structure) reduces the quark-quark channel to a single vector in color space.  Therefore condensation in this channel  can break color symmetry only partially, in the pattern $SU(3) \rightarrow SU(2)$.   Some gluons remain massless, and some quarks remain gapless, so infrared divergences remain.   

\subsubsection{Ground State}

Simplicity and self-consistency (that is, consistent use of weak coupling) first arrive when we consider three flavors.   

I'll describe the full structure of the condensate momentarily, but since that's a little intimidating let me begin with a sketch.   Since the spin (singlet) and spatial ($s$-wave) structures are unremarkable I'll suppress them, and also chirality.   The favored condensate should be antisymmetric in color and in flavor, which suggests the form
$$
\langle q^\alpha_a q^\beta_b \rangle \ \sim \  \epsilon^{\alpha \beta *}  \epsilon_{ab *}
$$
where the Greek indices are for color, the Latin indices are for flavor, and * is a wildcard.   Now by setting the wild cards equal, and contracting, we maintain as much residual symmetry as possible.   Any fixed choices for the wildcards will break both color and flavor symmetries.  But by {\it locking\/} color to flavor we maintain symmetry under the combined (so-called diagonal) symmetry group.   Thus we arrive at
\begin{equation}
\langle q^\alpha_a q^\beta_b \rangle \ \sim \  \epsilon^{\alpha \beta *}  \epsilon_{ab *} \ \rightarrow \ \epsilon^{\alpha \beta i}  \epsilon_{ab i} \ \propto (\delta^\alpha_a \delta^\beta_b - \delta^\alpha_b \delta^\beta_a)
\end{equation}
This condensate breaks local color times global flavor $SU(3) \times SU(3)$ to a diagonal,``modified flavor'' global $SU(3)$.   

It also spontaneously breaks baryon number symmetry.    To a particle physicist encountering these ideas for the first time, that might sound dramatic -- and it is, but not in the sense that it allows the material to decay.    With the sample enclosed in a finite volume, outside of which the order parameter vanishes, there is a strict conservation law for the integrated baryon number.   As in the theory of liquid helium 4, where one speaks of a condensate of helium atoms, the true implication is that there is easy transport of baryon number within the sample.   More specifically, there is a massless Nambu-Goldstone field, which supports the supercurrents characteristic of superfluidity.

Now comes the full structure, in all its glory:
\begin{eqnarray}
&\langle {\bf 1} | (q^\alpha_a)^i_L(\vec k) (q^\beta_b)^j_L(- \vec k) | {\bf 1} \rangle = \nonumber  \\
&\epsilon^{ij} \Bigl(v_1 (|\vec k |)  (\delta^\alpha_a \delta^\beta_b - \delta^\alpha_b \delta^\beta_a) + v_2 (|\vec k |) (\delta^\alpha_a \delta^\beta_b + \delta^\alpha_b \delta^\beta_a)\Bigr)  =  \nonumber \\ & - \Bigl(L \leftrightarrow R\Bigr).
\end{eqnarray}

Here some further words of explanation are in order.   The mid-Latin indices $i,j$ are for spin.  The ``$L$'' and ``$R$'' are for left and right chirality.  The relative sign between left and right condensates reflects conservation of parity. The functions $v_1 (|\vec k |),  v_2 (|\vec k |)$ are, for weak coupling, peaked near the Fermi surface.   Our preceding discussion anticipated the $v_1$ term, but the $v_2$ term is also allowed by the residual symmetry.  That latter term indeed emerges from calculations based on the microscopic theory, though with $v_1 >>  v_2$.    

Tracking chiral flavor symmetry and baryon number together with color, the implied breaking pattern is:
\begin{equation}
SU(3)_{\rm color} \times SU(3)_L \times SU(3)_R \times U(1)_B \ \rightarrow \ SU(3)_\Delta \times Z_2
\end{equation}
The residual $SU(3)_\Delta$ global symmetry, and the $Z_2$ of fermion (quark) number, can be used to classify the CFL state's low-energy excitations.    There is no residual local symmetry: All the color gluons have acquired mass.   A more refined analysis reveals that all the quarks have acquired gaps.

Finally, as a consequence of the underlying -- spontaneously broken -- baryon number and chiral symmetries we also have generalized ground states, obeying
\begin{eqnarray}
&\langle {\bf U}, \theta | (q^\alpha_a)^i_L(\vec k) (q^\beta_b)^j_L(- \vec k) | {\bf U}, \theta \rangle = \nonumber \\  
&\epsilon^{ij} e^{i\theta} \Bigl(v_1 (|\vec k |)  ({\bf U}^\alpha_a {\bf U}^\beta_b - {\bf U}^\alpha_b {\bf U}^\beta_a) + v_2 (|\vec k |) ({\bf U}^\alpha_a {\bf U}^\beta_b + {\bf U}^\alpha_b {\bf U}^\beta_a)\Bigr)  = \nonumber \\  &- \Bigl(L \leftrightarrow R\Bigr) 
\end{eqnarray}
for an any $SU(3)$ matrix {\bf U}.   These generalized ground states are related to one another by global baryon number (phase) or chiral transformations.   Low-frequency, long-wavelength modulation of the fields $\theta$ and {\bf U}, which represents slow motion within the vacuum manifold, generates the Nambu-Goldstone bosons.   

One more comment about the ground state is in order.   Throughout this discussion I've used the language of gauge symmetry breaking and gauge non-singlet order parameters.   This is quite familiar and traditional in BCS theory, and also in the standard model of electroweak interactions.   Strictly speaking, however, it based on a lie, for local gauge invariance is never broken.  Matrix elements of gauge-variant expectation values always vanish in the physical Hilbert space.   Indeed, the physical Hilbert space of a gauge theory is defined by restricting to gauge-invariant states.     The usual procedures of ``spontaneous symmetry breaking'' using gauge-variant operators are a tool --  a way of implementing favorable correlations in weak coupling.   Their physical content emerges when we use them as a calculational device, in weak coupling, to draw consequences for gauge-invariant quantities such as the physical spectrum or the expectation values of gauge-invariant operators.   In the CFL phase, we can identify two non-zero gauge invariant vacuum expectation values that break chiral or baryon-number symmetries.  They are
\begin{eqnarray}
&\langle q_L q_L \bar {q_R} \bar {q_R} \rangle \nonumber \\
&\langle qqqqqq \rangle
\end{eqnarray}
with the color indices suitably contracted.   These expectation values arise as powers of the primary, gauge-variant condensates.   (After including instanton effects, we also get $\langle q_L \bar{q_R} \rangle$.)    By way of contrast, neither conventional $s$-wave spin-singlet BCS condensation nor doublet condensation in the standard electroweak model support true order parameters.

\subsubsection{Elementary Excitations}

We can analyze the elementary excitations from the point of their spin and quantum numbers under the residual $SU(3)_\Delta$ symmetry.   There are three types:
\begin{enumerate}
\item {\it Excitations produced by the quark fields}:  They are spin-$\frac{1}{2}$ fermions that decompose as ${\bf 3} \times \bar{\bf 3} \rightarrow {\bf 8} + {\bf 1}$ under $SU(3)_{\rm color} \times SU(3)_{\rm flavor} \rightarrow SU(3)_\Delta$.   The singlet turns out, at weak coupling, to be significantly heavier than the octet.    
\item {\it Excitations produced by the gluon fields}: They are spin-1 bosons that form an octet.   
\item {\it Collective excitations}: They are a pseudoscalar octet of Nambu-Goldstone bosons, plus the singlet superfluid mode.  
\end{enumerate}

Overall, there is a striking resemblance between this calculated spectrum of low-lying excitations and what one might expect for the elementary excitations in the ``nuclear physics'' of QCD -- that is, the nuclear physics of QCD with three massless flavors -- based on standard concepts in QCD phenomenology and modeling.    The calculated elementary excitations map nicely onto the entries in the expected hadron spectrum.   Even the superfluid mode makes sense, because we would expect, in this idealized ``nuclear physics'',  pairing to occur in the dibaryon channel.    

Since conventional (heuristic) ``nuclear physics'' and the asymptotic (calculated) CFL state match so well with regard both to their ground state symmetry and to their low-lying spectrum, it is hard to avoid the conjecture that there is no phase transition separating these states.   Consider cranking up the chemical potential, starting from zero.   First there's Void.  At a critical value nuclear matter appears, with a first-order transition.  After that, there's just smooth evolution.   

This conjecture of {\it quark-hadron continuity\/} is both (superficially) paradoxical and conceptually powerful.   

The claim that the baryons of conventional ``nuclear physics'' are supposed to go over smoothly into excitations produced directly by single quark fields is paradoxical.   After all, baryons are famous for containing three quarks, and three can't evolve smoothly into one!    Well, actually it can.   When space is filled with a condensate of quark pairs, the difference between three and one is negotiable.    

Quark-hadron continuity is a powerful conceptual claim: It implies that the calculable forms of confinement and chiral symmetry breaking we construct by adapting the methods of BCS theory are in the same universality class as confinement and chiral symmetry breaking at low energies, within nuclear (or rather ``nuclear'') matter.

\subsubsection{Electric Charge}

Absence of long-range forces and massless gluons is a rather bloodless characterization of confinement.   What we'd really like to explain is: Why don't fractionally charged particles appear in the spectrum, given that they're in the Lagrangian?  

To address that question, we must couple electromagnetism into our theory.   Of course, electromagnetism is connected with the photon, which couples as
\begin{equation}
\gamma : e \left(\begin{array}{ccc}\frac{2}{3} & 0 & 0 \\0 & -\frac{1}{3} & 0 \\0 & 0 & -\frac{1}{3}\end{array}\right)
\end{equation}
to flavor indices, in an evident notation.   The symmetry associated with this generator is broken in the CFL ground state.   However there is a related gluon, which couples as 
\begin{equation}
\Gamma : g \left(\begin{array}{ccc}\frac{2}{3} & 0 & 0 \\0 & -\frac{1}{3} & 0 \\0 & 0 & -\frac{1}{3}\end{array}\right)
\end{equation}
to color indices.   The combination
\begin{equation}
\tilde \gamma \ = \ \frac{g\gamma + e \Gamma}{\sqrt {g^2 + e^2}}
\end{equation}
leaves the mixed Kronecker deltas that characterize the CFL ground state invariant, so it defines a massless gauge boson.   $\tilde \gamma$ is a modified photon, that defines the meaning of electromagnetism to an observer living within the CFL ground state.    Formally, for $g >> e$, it goes over into the ordinary photon.   However in that limit the small part $\Gamma$ couples much more strongly than the large part $\gamma$, and basic properties of the modified photon are, in fact, modified.   

Since the electron sees only $\gamma$, we read off its effective $\tilde \gamma$ charge as $-\frac{e}{\sqrt {g^2 + e^2}}$.   Excitations produced by quark fields get a contribution of either $\frac{2}{3} e \times \frac{g}{\sqrt {g^2 + e^2}}$ or  $-\frac{1}{3} e \times \frac{g}{\sqrt {g^2 + e^2}}$ from $\gamma$ and a contribution of either $-\frac{2}{3} g \times \frac{e}{\sqrt {g^2 + e^2}}$ or  $\frac{1}{3} g \times \frac{e}{\sqrt {g^2 + e^2}}$ from $\Gamma$.   Adding the two contributions, you see that the quarks are either neutral, or their total $\tilde \gamma$ charge is $\pm 1$ times the charge of the electron.    The possible charges of quarks in the CFL phase match the observed charges of baryons, which is a nice check on the quark-hadron continuity conjecture. Precisely these integer charge assignments appeared in the early work of Han and Nambu, who introduced color degrees of freedom together with a non-trivial embedding of electromagnetism to achieve them.     Similarly, the gluon and pseudoscalar meson charges are all integral (and match what you find in the particle data tables).     

A similar transmutation of the charge spectrum occurs in the standard model of electroweak interactions.  In that context, the $SU(2)$ gauge symmetry of weak isospin and the $U(1)$ gauge symmetry of hypercharge are separately broken, but a linear combination survives to become electromagnetism.   To reproduce the known electric charges of quarks, leptons, and $W$ bosons one must postulate a very peculiar spectrum of fractional hypercharges.     Unified theories of the strong, weak, and electromagnetic interactions, based on symmetry groups such as $SU(5)$ or $SO(10)$, which extend the standard model $SU(3)\times SU(2) \times U(1)$, arrive quite naturally at that very peculiar spectrum.  That is perhaps the most compelling evidence that such theories are on the right track \cite{unificationReview}.   

\subsubsection{Material Properties}

What happens to matter, if you keep squeezing?  Ultimately -- that is, for chemical potentials well above the strange quark mass but well below the charm quark mass -- it goes into the CFL phase, in which
\begin{itemize}
\item  Hadronic matter forms a transparent insulator.   We've discussed how the photon gets modified.   Some of the massless Nambu-Goldstone bosons are electrically charged, but once we take into account non-zero quark masses, these bosons (apart from the superfluid mode, which is electrically neutral) acquire mass.  Thus all the charged excitations have a gap, and we get an insulator.   Note especially that while it is  a {\it color\/} superconductor, hadronic mater in the CFL phase is not an {\it electrical\/} superconductor.   
\item It is a superfluid.   
\item It is vastly different from ordinary nuclear matter.   It contains an equal mix of strange quarks, for one thing, and strong trans-baryon correlations among the quarks.   One might expect, and model calculations tend to show, that there is a sharp transition between the two phases of hadronic matter, namely nuclear and CFL, including an abrupt jump in density.    
\end{itemize}
The first two items suggest that by squeezing we ultimately arrive at material similar to liquid helium 4, but of course with vastly higher density.    The third item suggests possibilities for astrophysical signatures.  (As I'll discuss momentarily, at present we cannot preclude the possibility of additional phases of hadronic at intermediate densities.)   I expect that eventually observation of gravitational waves from the final infall of neutron star - neutron star or neutron star - black hole binaries, in particular, will bring our knowledge of neutron star interiors to a new, much higher level.  Then predictions of this sort will be tested.

\subsection{Beyond Color-Flavor Locking}

As I  emphasized earlier, ordinary real-life nuclear matter is quite different from CFL.   In practice, the effect of the strange quark's mass on QCD phenomenology is far from negligible.  At low density the energetic cost of the strange quark outweighs its possible advantage in interaction energy, and ordinary nuclear matter has zero strangeness.   Because the condensation mechanism at the heart of CFL necessarily connects three different flavors, and must bring in strange quarks, it does not apply to ordinary nuclear matter.    CFL will set in when the chemical potential is sufficiently high that the strange quark mass is relatively negligible.   That will occur for asymptotically high chemical potentials -- or, equivalently, sufficiently high densities. Unfortunately, at present our calculational ability is not up to the task of predicting what happens subasymptotically.   It is possible that nuclear matter transitions directly to CFL; it is also possible that there are additional intermediate states.    Even if the transition is abrupt, as I suspect it is, presently we can't  predict the chemical potential at which it occurs, nor the jump in density that accompanies it.   These uncertainties hamstring our ability to make crisp astrophysical applications.  

BCS pairing works best when the modes being paired have close to zero (free) energy.  Ideally, many pairs should share the same quantum numbers, so that we can get enhancement factors from their coherent contributions. In that case, superconductivity can be triggered by arbitrarily weak interactions.    On the other hand if the Fermi surfaces of the quark species we'd like to pair don't match, so that modes at $\vec k$ and $-\vec k$ can't both be close to their respective Fermi surfaces simultaneously, then some compromise is necessary.  There are several possibilities:
\begin{itemize}
\item Meson condensates, involving the Nambu-Goldstone bosons, can soak up some of the unwanted flavor imbalance.
\item Less desirable forms of pairing, such as $p$-wave, that can work with a single flavor, might occur.   $p$-wave pairing (in three dimensions) leaves gapless fermions, which might themselves pair, forming a secondary condensate at a lower energy scale.   
\item Pairing can occur at one or several non-zero wave-vectors, i.e. involving modes $(\vec k + \vec \kappa, -\vec k)$.   These phases, which break translation invariance, are known as LOFF (Larkin-Ovchinnikov-Ferrell-Fulde) phases, or crystalline superconductivity.  
\item Pairing can occur between modes that are (nominally) particle and hole; i.e. one can dig into a Fermi ball, or supply a shell, to take advantage of potential energy gains at the cost of kinetic energy.   If a vestige of the original Fermi surface remains, one has a breached (gapless) phase, where a superfluid condensate coexists with a normal fluid component.
\end{itemize}

One bright spot is that cold atom physicists are beginning to explore traps loaded with several fermion species.   In that context it is totally natural to have Fermi surfaces of different sizes and couplings that are not small, so that similar complications arise, but now in systems that are experimentally accessible and allow controlled manipulation of the underlying parameters.   There has already been a fruitful cross-migration of ideas and information between these fields, and I expect that will continue.

\section{Gauge-Rotation Locking and Quantum Statistics: Anyons \cite{anyonReview}}

\subsection{Gauge-Rotation Locking}

Consider a $U(1)$ gauge theory that  is spontaneously broken by a condensate associated with a field of $\rho$ of  charge $mq$, with $m$ an integer.  Assume the theory also contains particles of charge $q$, associated with a field $\eta$, which does {\it not\/} condense.  Gauge transformations that multiply $\rho$ by $e^{2\pi i k}$will multiply $\eta$ by $e^{2\pi ik/m}$.  For integer $k$ such transformations leave the condensate invariant, but not necessarily $\eta$.   We are therefore left with an unbroken gauge group $Z_m$, the integers modulo $m$, that is not entirely trivial, at least mathematically.   On the other hand, no conventional long-range gauge interaction survives the symmetry breaking.  Does the residual symmetry have any physical consequences?   

Indeed it does.  They are subtle, but very interesting indeed.  
  
The theory supports vortices with flux quantized in units of
\begin{equation}
\Phi_0 \, = \, \frac{2\pi}{mq}
\end{equation}
(in units with $\hbar \equiv 1$).   

The flux is associated with a gauge potential, whose azimuthal piece we can take to have the form
\begin{eqnarray}
A_\theta (r, \theta) \  &=&  \frac{\Phi}{2\pi} f(r) \nonumber \\
f(0) \ &=& 0  \nonumber \\
f(\infty) &=& 1
\end{eqnarray}
and a condensate of the form
\begin{eqnarray}
\rho (r, \theta) \ &=& g(r) e^{i \frac{\Phi_{\ }}{\Phi_{0}}\theta} \nonumber \\
g(0) \ &=& 0 \nonumber  \\
g(\infty) &=& 1
\end{eqnarray}

This condensate is neither rotationally invariant nor gauge invariant.   It is, however, invariant under the combined rotation+gauge transformation
\begin{equation}
\tilde L \ = \ L + \Lambda \equiv -i\frac{\partial}{\partial \theta} - \frac{\Phi_{\ }}{\Phi_{0}} \frac{Q}{m}
\end{equation}
where $Q$ is the charge operator.  Here $\Lambda$ is a generator of spatially constant gauge transformations.    Thus there is a residual modified rotational symmetry, locking naive spatial rotations to appropriate gauge transformations, under which the vortex is invariant.   Indeed, from a strictly logical perspective it would be preferable to postulate the symmetry, and use it to motivate the vortex {\it ansatz}.   

For the condensate field $\rho$, which has charge $m$, the new contribution to the angular momentum is an integer.  Indeed, its ``role in life'' is to convert the spatial form of the condensation, which whirls in the partial wave with angular momentum $\frac{\Phi_{\ }}{\Phi_{0}}$, so that it represents, at spatial infinity, the state of rest.  More formally, the {\it kinetic\/} angular momentum, which is the gauge invariant version, gets annulled at infinity, by cancellation between the ordinary gradient and vector potential terms in the covariant derivative $D_\theta = \partial_\theta + i Q A_\theta$.    The square of angular momentum contributes to the energy density, so this cancellation must occur at spatial infinity, in order that the total energy of the vortex (in two dimensions), or the energy per unit length (in three dimensions) remains finite.   For broken global symmetries this cancellation is not an option.  In that case the vortices in that case have logarithmically divergent energy, and carry angular momentum; these facts underly the vastly different phenomenology associated with vortices in superconductors versus liquid helium.   

For the quanta of fields whose charge is not an integer multiple of the charge $Q$ of $\rho$, on the contrary, the new contribution to the angular momentum is {\it not\/} necessarily an integer, due to the factor $Q/m$.  So composites formed from particles of these kind and vortices will, in general, carry fractional angular momentum.    

Since one expects, on very general grounds, that there ought to be a tight connection between the spin of a particle and its quantum statistics, we are led to look into the quantum statistics of these objects, anticipating something unusual.

\subsection{Anyons}

Traditionally, the world has been divided between bosons (Bose-Einstein statistics) and fermions (Fermi-Dirac statistics).   Let's recall what these are, and why they appear to exhaust the possibilities.  

If two identical particles start at positions $(A, B)$ and transition to  $(A^\prime, B^\prime)$, we must consider both $(A, B) \rightarrow (A^\prime, B^\prime)$ and $(A, B) \rightarrow (B^\prime, A^\prime)$
as possible accounts of what has happened.  According the rules of quantum mechanics, we must add the amplitudes for these possibilities, with appropriate weights.   The rules for the weights encode the dynamics of the particular particles involved, and a large part of what we do in fundamental physics is to determine such rules and derive their consequences.   

In general, discovering the rules involves creative guesswork, guided by experiment.  One important guiding principle is correspondence with classical mechanics.  If we have a classical Lagrangian $L_{\rm cl.}$, we can use it, following Feynman, to construct a path integral, with each path weighted by a factor 
\begin{equation}\label{classicalWeight}
e^{i\int dt L_{\rm cl.}} \equiv e^{iS_{\rm cl.}}
\end{equation}
where $S_{\rm cl.}$ is the classical action.  This path integral provides -- modulo several technicalities and qualifications --  amplitudes that automatically implement the general rules of quantum mechanics.  Specifically: it sums over alternative histories, takes products of amplitudes for successive events, and generates unitary time evolution.   

The classical correspondence, however, does not instruct us regarding the relative weights for trajectories that are topologically distinct, i.e. trajectories that cannot be continuously deformed into one another.   Since only small variations in trajectories are involved in determining the classical equations of motion,  from the condition that $S_{\rm cl.}$ is stationary, the classical equations cannot tell us how to interpolate between topologically distinct trajectories.  We need additional, essentially quantum-mechanical rules for that.   

Now trajectories that transition $(A, B) \rightarrow (A^\prime, B^\prime)$ respectively $(A, B) \rightarrow (B^\prime, A^\prime)$ are obviously topologically distinct.   The traditional additional rule is: For bosons, add the amplitudes for these two classes of trajectories\footnote{As determined by the classical correspondence, or other knowledge of the interactions.}; for fermions, subtract.  

Those might appear to be the only two possibilities, according to the following (not-quite-right) argument.  Let us focus on the case  $A = A^\prime, B = B^\prime$.   If we run an ``exchange'' trajectory $(A, B) \rightarrow (B, A)$ twice in succession, the doubled trajectory is a direct trajectory.   The the square of the factor we assign to the exchange trajectory must be the square of the (trivial) factor $1$ we associate to the direct trajectory, i.e. it must be $\pm 1$.  

The argument in the preceding paragraph is not conclusive, however, because there can be additional topological distinctions among trajectories, not captured by the permutation among endpoints. This distinction is especially important in 2 spatial dimensions, so let us start there.   (I should recall that quantum-mechanical systems at low energy can effectively embody reduced dimensionality, if their dynamics is constrained below an energy gap to exclude excited states whose wave functions have structure in the transverse direction.)  

 The topology of trajectory space is then specified by the {\it braid group}.   Suppose that we have $N$ identical particles.  Define the elementary operation $\sigma_j$ to be the act of taking particle $j$ over particle $j+1$, so that their final positions are interchanged, while leaving the other particles in place.  (See Figure 1.)  We define products of the elementary operations by performing them sequentially.  Then we have the obvious relation 
\begin{equation}\label{distantCommutation}
\sigma_j \sigma_k \, = \, \sigma_k \sigma_j; \, \, \, \, |j-k| \geq 2
\end{equation}
among operations that involve separate pairs of particles.   
We also have the less obvious {\it Yang-Baxter\/} relation
\begin{equation}\label{YangBaxter}
\sigma_j \sigma_{j+1} \sigma_j \, = \, \sigma_{j+1} \sigma_j \sigma_{j+1}
\end{equation}
which is illustrated in Figure 1.  The topologically distinct classes of trajectories are constructed by taking products of $\sigma_j$s and their inverses, subject only to these relations.

In three dimensions there is more room to maneuver the strands, and we have an additional relation 
\begin{equation} \label{squareTriviality}
\sigma_j^2 \, = \, 1
\end{equation}
When these relations are added to the previous ones, we find that the braid group reduces to the ordinary so-called symmetric group $S_N$ of permutations on $N$ letters.   An interesting intermediate possibility is to demand the relation that rotations through $4\pi$ are trivial but rotations through $2 \pi$ might not be, as in the mathematics of spinors.   Then one would impose
\begin{equation} \label{quadTriviality}
\sigma_j^4 \, = \, 1
\end{equation}
in place of Eqn. (\ref{squareTriviality}).    This distinction will come up again shortly.     

%\begin{figure}[ht]\label{figure1}
%\epsfxsize=8cm
%\centerline{\epsfbox{figures1.pdf}}
%\caption{The elementary acts of crossing one particle trajectory over another generate the braid group.    The Yang-Baxter relation $\sigma_1 \sigma_{2} \sigma_1 = \sigma_{2}\sigma_1 \sigma_{2}$, made visible here, is its characteristic constraint.}
%\end{figure}

\begin{figure}[ht]\label{figure1}
%\epsfxsize=8cm
%\centerline{\epsfbox{figures1.pdf}}
\includegraphics[width=8cm]{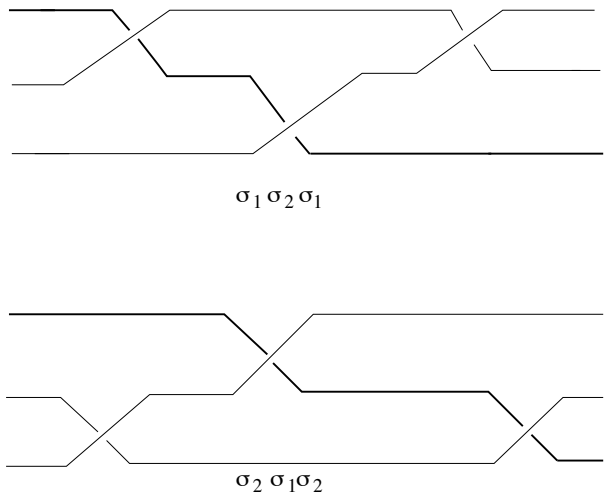}
\caption{The elementary acts of crossing one particle trajectory over another generate the braid group.    The Yang-Baxter relation $\sigma_1 \sigma_{2} \sigma_1 = \sigma_{2}\sigma_1 \sigma_{2}$, made visible here, is its characteristic constraint.  The top configuration can slide smoothly into the bottom one, with endpoints held fixed.}
\end{figure}

The defining equations Eqn. (\ref{distantCommutation}, \ref{YangBaxter}) for the braid group allow a continuous range of 1-dimensional unitary representations, of the very simple form
\begin{equation}
\sigma_j = e^{i\theta}
\end{equation}
for all $j$, with $\theta$ an arbitrary real number.   One can have any phase, not just the $\pm 1$ characteristic of bosons and fermions.   For that reason I christened particles carrying more general quantum statistics {\it anyons}. 

With this very general background in mind, let us return to our fractional angular momentum vortices.
A particle or group of particles with charge $bq$ moving around a flux $\Phi$ acquires, according the minimal coupling gauge Lagrangian, phase
\begin{equation}
\exp ibq(\oint dt \vec v \cdot \vec A)  \, = \,  \exp ibq(\oint d\vec x  \cdot \vec A ) \, = \, e^{i \Phi bq} 
\end{equation}
(Note that in two dimensions the familiar flux tubes of three-dimensional physics degenerate to points, so it is proper to regard them as particles.)  If the flux is $a\Phi_0$, then the phase will be $e^{2\pi i \frac{ab}{m}}$.   Note that this phase does not depend on the velocity, curvature, or any details of the particles' dynamics, other than the topology of how their world-lines interweave.  For that reason, we say we have a topological interaction.   

Composites with $(\rm{flux}, \rm{charge})$ $ = (a \Phi_0, bq)$ will be generally be particles with unusual quantum statistics.  For as we implement the interchange $\sigma_j$, each charge cluster feels the influence of the other's flux, and non-trivial phase is required.  A close analysis shows that the anomalous statistics is just such as to preserve a spin-statistics connection, in the form
\begin{equation}
e^{2\pi i J} \ = \ e^{i\theta}
\end{equation}

Evidently quantum statistics, both conventional and unconventional, can be regarded as a special type of long-range interaction.  It is remarkable that this interaction is not associated with the exchange of any massless particle.  Indeed, our specific model, with broken gauge symmetry, can be fully gapped.    One can also have topological interactions, involving similar accumulations of phase, for non-identical particles.  What governs these topological interactions are the quantum numbers, or more formally the superselection sectors, of the particles, excitations, or clusters involved, not their detailed internal structure.

The phase factors that accompany winding have observable consequences.   They lead to a characteristic ``long range'' contribution to the scattering cross-section, specifically 
\begin{equation}
\frac{d\sigma}{d\phi} \ = \ \sin^2 (\pi \frac{ab}{m})    \frac{1}{2\pi k} \frac{1}{\sin^2{\frac{\phi}{2}} }
\end{equation}
It diverges at small momentum transfer and in the forward direction. A cross section of this kind was first computed by Aharonov and B\"ohm \cite{AharonovB} in their classic paper on the significance of the vector potential in quantum mechanics.   If we could do experiments in the style of high-energy physics, forming beams of quasiparticles and scattering them, we'd be in great shape.  Unfortunately, as a practical matter the highly characteristic cross-sections associated with anyons may not be easy to access experimentally for the examples that occur as excitations in exotic states of condensed matter (although it could be worth a try!).   

%\begin{figure}[ht]\label{figure2}
%\epsfxsize=10cm
%{
%%\epsfbox{figure21bis.pdf}
%\epsfbox{figure4bis.pdf}}
%%{\epsfbox{figure23bis.pdf}
%\caption{A schematic interference experiment to reveal quantum statistics.   We study how the combined current depends on the occupation of the quasiparticle island.}
%\end{figure}
\begin{figure}[ht]\label{figure2}
%\epsfbox{figure21bis.pdf}
\includegraphics[width=10cm]{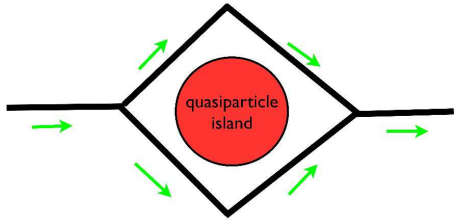}
%{\epsfbox{figure23bis.pdf}
\caption{A schematic interference experiment to reveal quantum statistics.   One measures how the combined current depends on the occupation of the quasiparticle island.}
\end{figure}

Interferometry appears more practical.    The basic concept is simple and familiar, both from optics and (for instance) from SQUID magnetometers.  One divides a coherent flow into two streams, which follow different paths before recombining.   The relative phase between the paths determines the form of the interference, which can range from constructive to destructive recombination of the currents.   We can vary the superselection sector of the area bounded by the paths, and look for corresponding, characteristic changes in the interference.  (See Figure 2.)   Though there are many additional refinements, this is the basic concept behind both Goldman's suggestive experiments \cite{goldman} and other planned anyon detection experiments \cite{plannedExpts}.

Elementary excitations in the fractional quantum Hall effect are predicted to be anyons.    A proper discussion of that field would require a major digression, which would not be appropriate here.  For the central calculation see \cite{qHallAnyons}, for extensive discussion and review, see \cite{anyonReview}.     I'd like to emphasize, in any case, that the general concept of anyons is by no means restricted to the quantum Hall effect; on the contrary, I believe the subject will reach a new level of interest and importance as more robust, user-friendly realizations are discovered.   This is a most important area for future research.

%I'll say more about that below. %By far the simplest states to analyze are the original Laughlin $1/m$ states, where the excitations are anyons with $\theta = \pi /m $.   There is a rich theory covering more general cases.    

\subsection{Nonabelian Anyons}

The preceding field-theoretic setting for abelian anyons immediately invites nonabelian generalization.   We can have a nonabelian gauge theory broken down to a discrete nonabelian subgroup; vortex-charge composites will then exhibit long range, topological interactions of the same kind as we found in the abelian case, for the same reason.   

Though the starting point is virtually identical, when we consider interactions among several anyons the mathematics and physics of the nonabelian case quickly becomes considerably more complicated than the abelian case, and includes several qualitatively new effects.   First, and most profoundly, we will find ourselves dealing with irreducible {\it multidimensional\/} representations of the braiding operations.   Thus by winding well-separated particles\footnote{From here on I will refer to the excitations simply as particles, though they may be complex collective excitations in terms of the underlying electrons, or other degrees of freedom.}  around one another, in principle arbitrarily slowly, we can not only acquire phase, but even navigate around a multidimensional Hilbert space.   For configurations involving several well-separated particles, the size of the many-body ``ground state'' Hilbert spaces can get quite large: roughly speaking, they grow exponentially in the number of particles.   Since all the states in this Hilbert space are related by locally trivial -- but globally non-trivial -- gauge transformations, they should be very nearly degenerate.   This situation is reminiscent of what one would have if the particles had an internal of freedom -- a spin, say.   However here the emergent degrees of freedom here are not localized on the particles, but more subtle and globally distributed.    

The prospect of contructing very large Hilbert spaces that we can navigate in a controlled way using topologically defined (and thus forgiving!), gentle operations in physical space, and whose states differ in global properties not easily obscured by local perturbations, has inspired visions of {\it topological quantum computing}.   (Preskill \cite{preskill} has written an excellent introductory review.)  The journey from this exalted vision to  real-world engineering practice will be challenging, to say the least, but thankfully there are fascinating prospects along the way.   

%\begin{figure}[ht]\label{figure3}
%\epsfxsize=5cm
%{
%\epsfbox{figure21bis.jpg}
%\epsfbox{figure22bis.jpg}}
%{\epsfbox{figure23bis.pdf}
%%\epsfbox{figures3.pdf}
%}
%\caption{By a gauge transformation, the vector potential emanating from a flux point can be bundled into a singular line.   This aids in visualizing the effects of particle interchanges.   Here we see how nonabelian fluxes, as measured by their action on standardized particle trajectories, are modified by particle interchange.}
%\end{figure}
\begin{figure}[ht]\label{figure3}
\includegraphics[width=11cm]{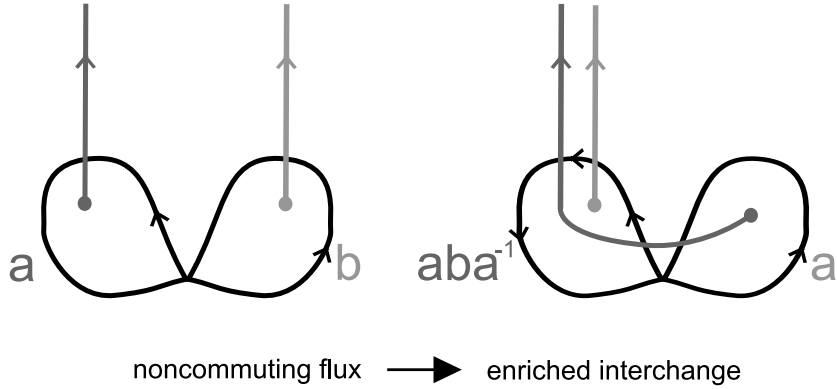}
%\epsfbox{figures3.pdf}
\caption{By a gauge transformation, the vector potential emanating from a flux point can be bundled into a singular line.   This aids in visualizing the effects of particle interchanges.   Here we see how nonabelian fluxes, as measured by their action on standardized particle trajectories, are modified by particle interchange.}
\end{figure}

The tiny seed from which all this complexity grows is the phenomenon displayed in Figure 3.  To keep track of the topological interactions, it is sufficient to know the total (ordered) line integral of the vector potential around simple circuits issuing from a fixed base point.   This will tell us the group element $a$ that will be applied to a charged particle as it traverses that loop.   (The value of $a$ generally depends on the base point and on the topology of how the loop winds around the regions where flux is concentrated, but not on other details.   More formally, it gives a representation of the fundamental group of the plane with punctures.)   If a charge that belongs to the representation $R$ traverses the loop, it will be transformed according to $R(a)$.   With these understandings, what Figure 3 makes clear is that when two flux points with flux $(a, b)$ get interchanged by winding the second over the first, the new configuration is characterized as $(aba^{-1}, a)$.    Note here that we cannot simply pull the ``Dirac strings'' where flux is taken off through one another, since nonabelian gauge fields self-interact!    So motion of flux tubes in physical space generates non-trivial motion in group space, and thus in the Hilbert space of states with group-theoretic labels.   

%\begin{figure}[ht]\label{figure4}
%\epsfxsize=8cm
%\centerline{\epsfbox{figures3.pdf}}
%\caption{Winding a flux-antiflux pair around a text flux, and seeing that it gets conjugated, we learn find that charge the pair carries charge.}
%\end{figure}
\begin{figure}[ht]\label{figure4}
\includegraphics[width=10cm]{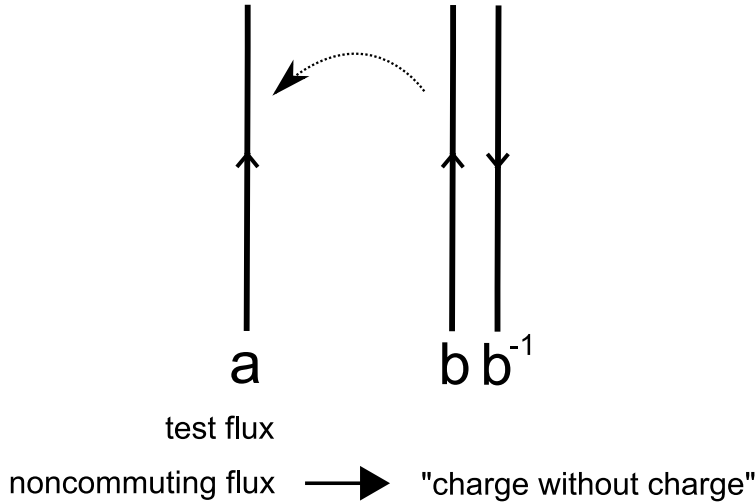}
\caption{Winding a flux-antiflux pair around a test flux, and seeing that its elements get conjugated, we learn that the pair generally carries charge.}
\end{figure}	

As a small taste of the interesting things that occur, consider the slightly more complicated situation displayed in Figure 4, with a pair of fluxes $(b, b^{-1})$ on the right.   It's a fun exercise to apply the rule for looping repeatedly, to find out what happens when we take this pair all the way around $a$ on the right.   One finds
\begin{equation}
(a, (b, b^{-1})) \,  \rightarrow \, (a, (aba^{-1}, a b^{-1} a^{-1}))
\end{equation}
i.e., the pair generally has turned into a different (conjugated) pair.   Iterating, we eventually close on a finite-dimensional space of different kinds of pairs.    There is a non-trival transformation ${\tilde R}(a)$ in this space that implements the effect of the flux  $a$ on pairs that wind around it.  But this property -- to be transformed by the group operation -- is  the defining property of charge!   We conclude that flux pairs -- flux and inverse flux --  act as charges.  We have constructed, as John Wheeler might have said, Charge Without Charge.   

Its abstract realization through flux tubes makes it manifest that nonabelian statistics is consistent with all the general principles of quantum field theory.   Practical physical realization in condensed matter is a different issue, for in that context, nonabelian gauge fields aren't ready to hand.   

Fortunately, and remarkably, there may be other ways to get there.  At least one state of the quantum Hall effect, the so-called Moore-Read state at filling fraction $\frac{5}{2}$, has been identified as a likely candidate to support excitations with nonabelian statistics \cite{mooreRead}.    

The nonabelian statistics of the Moore-Read state is closely tied up with spinors \cite{nayakFW} \cite{ivanov}.   I'll give a proper discussion of this, including an extension to 3 + 1 dimensions, elsewhere \cite{halflings}.   Here, I'll just skip to the chase.   Taking $N$ $\gamma_j$ matrices satisfying the usual Clifford algebra relations
\begin{equation}
\{ \gamma_j , \gamma_k \} \, = \, 2\delta_{jk}
\end{equation}
the braiding $\sigma_j$ are realized as
\begin{equation}\label{cliffordBraids}
\sigma_j \, = \, e^{i\pi / 4} \frac{1}{\sqrt 2} (1 + \gamma_j \gamma_{j+1})
\end{equation}
It's an easy exercise to show that these obey Eqns. (\ref{distantCommutation}, \ref{YangBaxter}), and 
$\sigma_j^4 =1$ (Eqn. (\ref{quadTriviality})) but not $\sigma_j^2 =1$ (Eqn. (\ref{squareTriviality})).  

\subsection{Pairing, Statistical Transmutation, and Zero Modes}

Finally, it's appropriate to mention that there's a deep connection between BCS theory and the sorts of quantum Hall states that support nonabelian anyons.   They are connected adiabatically -- in a conceptual parameter space -- through {\it statistical transmutation}.     

The imposition of a constant magnetic field on an two-dimensional electron gas is not a uniformly small perturbation, even if the magnitude of the field $B$ is tiny.   That is because in formulating quantum mechanics the Hamiltonian is fundamental, and in the Hamiltonian the vector potential $\vec A$ appears.   Stokes' law informs us that the vector potential associated with a constant field strength grows linearly with the size of the sample.  Thus large quantities appear in Hamiltonian, and the perturbation associated with a tiny magnetic field is not uniformly small.   And indeed we know that such a perturbation can induce a qualitative change in the spectrum, changing (say) the conventional parabolic free-electron spectrum into the quantized Landau levels, which feature a gap above a highly degenerate ground state.   

The idea of statistical transmutation is that we can cancel off the growing part of the magnetic vector potential, if we associate to each electron an appropriate change in quantum statistics.  Indeed, as I've reviewed above, one can effectively implement changes in the quantum statistics of particles by attaching notional flux and charge to those particles.   (In the early days I called this ``fictitious flux'', to distinguish it from electromagnetic flux.)   Now if we add up the notional gauge potentials from a constant density of electrons, we'll get -- again according to Stokes' law -- notional gauge potentials that grow linearly with the distance.    If we add the right amount of flux -- in other words, if we make a judicious change in quantum statistics -- we can arrange to make a cancellation between the parts of the real and notional gauge potentials which grow with distance.   The perturbation implementing this combined operation -- a small magnetic field applied {\it together with\/}  an appropriate small change in quantum statistics -- will then be uniformly small.    

The required relation between field and statistics can be neatly expressed as a connection between the filling fraction 
\begin{equation}
\nu \, \equiv \, \frac {\rho}{eB} \, , 
\end{equation}
where $\rho$ is the electron number density, and the quantum statistics parameter $\theta$.   We require
\begin{equation}\label{statTrans}
\Delta \frac{1}{\nu} \, = \, \Delta \frac{\theta}{\pi}
\end{equation}
Since gapped systems which lie along the lines defined by Eqn. (\ref{statTrans}) are related by a sequence of infinitesimal perturbations, we can expect that they lie in the same universality class, and will share universal properties.   

$\Delta {\theta} = 2\pi$ corresponds, on the one hand, to no net change in statistics, and on the other, to $\Delta \frac{1}{\nu} = 2$.   In this way our ``notional'' adiabatic path through anyons can connect proper (fermionic) electron states.    A notable example: the fractional Laughlin states at $\frac{1}{\nu } = 2m + 1$ can be connected adiabatically to the integer quantum Hall state at $\nu =1$ (in other words, $m=0$).    To my mind, this is the most profound\footnote{and the most under-appreciated ...} way to understand the existence of gapped many-body states at those filling fractions, and their other most distinctive properties.   

$\frac {1}{\nu} = 0$ corresponds to zero magnetic field.   In zero magnetic field, for appropriate attractive interactions, electrons can form a gapped superconductor, specifically a $p_x + ip_y$ superconductor, through BCS pairing in the $l=1$ channel.   According to the Eqn. (\ref{statTrans}), that BCS superconductor can be adiabatically connected to $\nu = \frac{1}{2}$ quantum Hall states, which should share its universal properties.   (The observed $\nu  = \frac{5}{2}$ state plausibly contains 2 inert Landau levels, so its active dynamics involves $\nu = \frac{1}{2}$.)  Prominent among the universal properties we can calculate in the BCS state are: the existence of a gap; the existence of neutral `pair-breaking' excitations; and the existence Majorana zero modes on vortices, leading to nonabelian statistics for those vortices.  The nonabelian statistics that arises here is of the kind I sketched earlier, in Eqn. (\ref{cliffordBraids}).  The Clifford algebra is realized here, concretely, as the algebra of the operator coefficients that multiply the vortex-centered zero modes in the expansion of the electron field.    All the aforementioned features should carry over into appropriate $\nu = \frac{1}{2}$ states.

\end{document}